\documentclass[conference
]{IEEEtran}
\IEEEoverridecommandlockouts    
\usepackage{verbatim} 
\usepackage{cite}
\usepackage{array}
\usepackage{bm}
\usepackage{amsmath,amssymb,amsfonts}
\usepackage{algorithmic}
\usepackage{graphicx}
\usepackage{tikz}
\usepackage{circuitikz}
\usepackage{siunitx}
\usepackage{gensymb}
\usetikzlibrary{shapes.geometric, arrows,chains, positioning}
\usetikzlibrary{arrows.meta} 
\usetikzlibrary { decorations.pathmorphing, decorations.pathreplacing, decorations.shapes, } 
\usepackage{pbox}
\usepackage{relsize}
\tikzset{
    pin/.style = {font = \relsize{-2}} 
}
\ctikzset{
    bipoles/length = 4em, 
    font = \relsize{-1}, 
}
\usepackage{textcomp}
\usepackage{xcolor}
\def\BibTeX{{\rm B\kern-.05em{\sc i\kern-.025em b}\kern-.08em
    T\kern-.1667em\lower.7ex\hbox{E}\kern-.125emX}}
\usepackage{mathtools}
\usepackage{schemabloc}
\usetikzlibrary{circuits}
\usetikzlibrary{arrows}
\makeatletter
\let\MYcaption\@makecaption
\makeatother
\usepackage[font=footnotesize]{subcaption}
\makeatletter
\let\@makecaption\MYcaption
\makeatother
\usepackage{caption}

\usepackage{pgfplotstable}
\usepackage{multirow}
\usepackage{mathtools}
\usepackage{commath}

\newcommand{\RNum}[1]{\uppercase\expandafter{\romannumeral #1\relax}}
\graphicspath{ {./confimages/} }
 \usepackage{pgfplots}
  \pgfplotsset{compat=newest}
  \usetikzlibrary{plotmarks}
  \usetikzlibrary{arrows.meta}
  \usepgfplotslibrary{patchplots}
  \usepackage{grffile}
  \usepackage{amsmath}
  
\usepackage{amsthm}

\usepackage{fancyhdr,graphicx,amsmath,amssymb}
\usepackage[ruled,vlined]{algorithm2e}
\usepackage{times}
\usepackage{hhline}

\usepackage{hhline}
\usepackage{booktabs} 
\usepackage{circuitikz}
\usepackage{siunitx}
\usepackage[inline]{enumitem}

\usepackage[hyphens]{url}
\usepackage{balance}
\usepackage[commandnameprefix=always]{changes}
\definechangesauthor[color=blue,name=Vassilis]{VP}
\usepackage{nicematrix}

\usepackage{flushend}



\begin{document}

\allowdisplaybreaks 

\title{Evaluation of RIS-Enabled B5G/6G Indoor Positioning and Mapping using Ray Tracing Models\vspace{-0.7cm}} 

\author{\IEEEauthorblockN{ 
Dimitris Kompostiotis\IEEEauthorrefmark{2}, Dimitris Vordonis\IEEEauthorrefmark{2}, Vassilis Paliouras\IEEEauthorrefmark{2}, George C. Alexandropoulos\IEEEauthorrefmark{5}, and Florin Grec\IEEEauthorrefmark{4}}
\IEEEauthorblockA{\IEEEauthorrefmark{2}Electrical and Computer Engineering Department, University of Patras, Greece,\\
 \IEEEauthorrefmark{5}Department of Informatics and Telecommunications, National and Kapodistrian University of Athens, Greece,\\
\IEEEauthorrefmark{4} European Space Agency (ESA), Noordwijk, Netherlands\\
e-mails: \{d.kompostiotis, d.vordonis\}@ac.upatras.gr, paliuras@ece.upatras.gr, alexandg@di.uoa.gr, Florin-Catalin.Grec@esa.int}}

\markboth{NOTES}%
{Shell \MakeLowercase{\textit{et al.}}: Bare Demo of IEEEtran.cls for IEEE Journals}

\maketitle
\begin{abstract}
A Reconfigurable Intelligent Surface (RIS) can significantly enhance network positioning and mapping, acting as an additional anchor point in the reference system and improving signal strength and measurement diversity through the generation of favorable scattering conditions and virtual line-of-sight paths.
In this paper, we present a comprehensive framework aimed at user localization and scatterer position estimation in an indoor environment with multipath effects. Our approach leverages beam sweeping through codebook-based beamforming at an $1$-bit RIS to scan the environment, applies signal component extraction mechanisms,
and utilizes a super-resolution algorithm
for angle-based positioning of both connected users and scatterers.
To validate the system's effectiveness, accurate 3D ray tracing models are employed, ensuring the robustness and effectiveness of the proposed approach in practical scenarios.
\end{abstract}

\begin{IEEEkeywords}
Reconfigurable intelligent surfaces, MIMO, positioning, localization, mapping, MUSIC, ray tracing. 
\end{IEEEkeywords}

\section{Introduction}
\label{sec:Intro}
The evolution of wireless networks has led to the development of new Internet of Things (IoT) services and applications, including smart cities, autonomous vehicles, transforming healthcare, process automation, and extended reality~\cite{wymeersch2022localisation,3gppTS261,3gppTS872,RIS_smart_cities}. 
These applications demand precise localization of both active network nodes and passive elements, like obstacles and scatterers. 
Achieving such high performance localization and mapping 
requires 
sufficient infrastructure coverage, the ability to resolve different signal paths, and accurate estimation of geometric parameters for each identifiable path. Currently, 
this estimation problem is solved using various radio technologies, such as Global Navigation Satellite Systems (GNSS)~\cite{teunissen2017springer,wymeersch2022radio} and Ultra-WideBand (UWB)~\cite{wymeersch2022radio}.
However, since GNSS is restricted to outdoor usage and UWB to short-range applications, new technologies are anticipated to bridge this gap by offering enhanced capabilities across different environments~\cite{wymeersch2022radio}.

Starting from 4-th Generation (4G) networks, localization is achieved using multiple anchors, such as Access Points (APs), which exchange signals with User Equipment (UE) to determine the Time Difference of Arrival (TDoA)~\cite{nikonowicz2024indoor} and pinpoint UE locations. In 5G networks, the use of multiple antennas at both APs and UEs has enabled the estimation of the Angles of Arrival (AoAs) and Angles of Departure (AoDs), facilitating environmental sensing and mapping~\cite{schmidt1986multiple,roy1989esprit,AVW22a}. Despite these advancements, current techniques face limitations in accuracy and reliability, particularly in multipath, interference-heavy, and Non-Line-of-Sight (NLoS) environments. 
To overcome some of these challenges, the use of Reconfigurable Intelligent Surfaces (RISs) has been proposed for 6G~\cite{alexandg_2021,Alexandropoulos2022Pervasive,basar2023RIS_mag}. RISs can enhance localization by providing higher angular resolution through their large physical apertures, and by addressing NLoS limitations when optimally placed~\cite{9721205,9847080}. By controlling the wireless propagation environment~\cite{kompostiotis2023received,kompostiotis2023secrecy,vordonis2022reconfigurable}, RISs facilitate the extraction of accurate channel parameter estimation, making the localization and mapping problem over-determined, and thus, easier to solve~\cite{9721205,RIS_loc}.

The potential of RIS-aided indoor localization and mapping is lately attracting various research investigations~\cite{9721205,RIS_loc,9847080,wymeersch2020radio,de2021convergent,chen2022reconfigurable,RIS_SLAM,RIS_compressed_ICASSP,RIS_drone_PIMRC}. RIS optimization is crucial in this field and various methods are being explored.  Approaches include RIS phase optimization tailored for fingerprinting solutions based on Received Signal Strength (RSS) measurements at the UE~\cite{zhang2021metalocalization}, and beam sweeping or hierarchical codebook strategies to scan the environment~\cite{albanese2021papir,he2020adaptive,RIS_low_overhead,RIS_hierarchical}. Furthermore, techniques for separating direct components from those generated by RIS reflections, as presented in~\cite{dardari2021nlos}, are vital for indoor and sub-6 GHz scenarios. Regarding the problem of mapping, Zhang \emph{et al.} \cite{zhang2024user} proposed a novel approach for scatterer position estimation using a MUltiple SIgnal Classification (MUSIC) based algorithm. Taha \emph{et al.} \cite{taha2023reconfigurable} introduced an innovative framework for estimating the depth map of the surrounding environment using a monostatic RIS-aided wireless sensing system, as well as an RIS codebook to create a sensing grid of reflected beams.
In addition, Tong \emph{et al.} \cite{tong2021joint} discretized the space and estimated the channel to identify scatterers in each sub-space, facilitating accurate environment estimation.

In this paper, a protocol for RIS-aided UE and scatterer positioning is presented. The proposed approach integrates codebook-based beamforming at the RIS, combined with a geometry-based method that facilitates angle-based positioning. Also, signal component extraction mechanisms are derived to distinguish direct signals from those received from reflections at a scatterer, and final results are validated using 3D Ray Tracing (RT) channel models. Unlike~\cite{zhang2024user}, which overlooks the RIS node in UE positioning and relies on the LoS conditions between the AP and UE, this work utilizes the RIS node and its reflection properties to compensate for the absence of the direct LoS AP-UE path. In addition,~\cite{zhang2024user} assumed continuous phase shifters at the RIS, whereas this work employs a practical and low-resolution $1$-bit RIS~\cite{RIS_1bit}; making the positioning problem more realistic and challenging.

The remainder of the paper is organized as follows: Section~\ref{s:systemmodel} describes the system model, while Sections~\ref{s:problem_localization} and~\ref{s:problem_mapping} present the protocol and the proposed estimation method for user and scatterer positioning, respectively. Section~\ref{s:simulations} includes the simulation results and Section~\ref{s:conclusion} concludes the paper.

\section{System and Signals Model}
\label{s:systemmodel}
\begin{figure}[t]
    \centering
    \scalebox{0.8}{\includegraphics{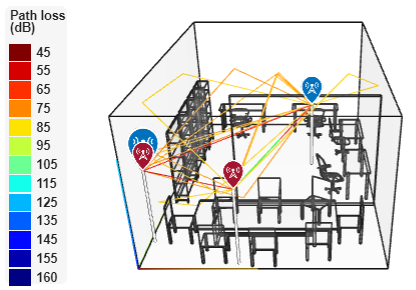}}
    \caption{The considered RT-based simulation setup where the red point denotes Tx positions, blue point the Rx, and the red and blue point the RIS node.}
    \label{fig:sim_setup1}
\end{figure}

This work considers the downlink RIS-enabled indoor wireless system shown in Fig.~\ref{fig:sim_setup1}, consisting of an AP equipped with $N_T$ antennas that communicates with an $N_{R}$-antenna Uniform Linear Array (ULA) UE. The direct channel paths from the AP to the UE are assumed as NLoS. The RIS is placed in a way that creates a LoS path, both at the AP-to-RIS, $\bm{H}_{1}[n]{\in} \mathbb{C}^{N \times N_T}$, and RIS-to-UE, $\bm{H}_{2}[n]{\in} \mathbb{C}^{N_R \times N}$, channels; $n$ denotes the specific time-domain channel filter tap. We model the RIS as a Unifom Rectangular Array (URA) transceiver and assume that it consists of $N {=} N_H {\times} N_V$ reflecting elements, where $N_H$ is the horizontal and $N_V$ is the vertical dimension. 
The direct AP-to-UE channel is represented by $\bm{H}_{d}[n] {\in} \mathbb{C}^{N_R \times N_T}$. To model all latter Multiple-Input and Multiple-Output (MIMO) wireless channels, the RT model of Matlab$\textsuperscript{TM}$ is adopted, which is a physics-based geometric wireless channel model and is ideal for describing static wireless channels in the frequency range of $100$~MHz--$100$~GHz. In discrete-time baseband representation, each channel is modeled for a specific carrier frequency $f_c$ using RT as follows:
\begin{equation}
\bm{H}[n] = \sum_{l=1}^{L}\beta_{l,n}\,\alpha_l\,e^{-\text{j}\,2\pi\tau_lf_c}\,\bm{\mathbf{a}}_{\text{Rx}}(\varphi_{\text{AoA}}^{l})\bm{\mathbf{a}}_{\text{Tx}}^{H}(\varphi_{\text{AoD}}^{l}),
\label{eq:RT-based_channel_model}
\end{equation}
where $L$ is the number of channel propagation paths, and for each $l$-th path, $\beta_{l,n} \in \mathbb{R}$ is the pathloss factor modeling the fractional-delay channel filter, $\alpha_l \in \mathbb{R}$ is the pathloss factor, $\tau_l$ is the propagation path delay, and $\varphi_{\text{AoA}}^{l}$ and $\varphi_{\text{AoD}}^{l}$ are the AoA and AoD, respectively. Finally, $\bm{\mathbf{a}}_{q}$ ($q{\in}\{\text{Tx},\,\text{Rx}\}$) is the array response vector which is given for the far-field case by:
\begin{equation}
    \mathbf{a}_{q}(\varphi, \theta) = \Big[e^{-\,\jmath\mathbf{k}(\varphi, \theta)^\mathsf{T}\mathbf{u}^{q}_1}, \ldots, e^{-\,\jmath\mathbf{k}(\varphi, \theta)^\mathsf{T}\mathbf{u}^{q}_N}\Big]^{\mathrlap{T}},
    \label{e:array_response_vec}
\end{equation}
with $\mathbf{u}^{q}_i$ being the Tx's ($q{\equiv} \text{Tx}$) or Rx's ($q{\equiv} \text{Rx}$) $i$-th transceiving element position and the wave vector ($\mathbf{k}(\varphi, \theta) {\in} \mathbb{R}^{3{\times} 1}$) of the transmitted signal propagating at the azimuth angle $\varphi$ and elevation $\theta$ with wavelength $\lambda$ is given by:
\begin{equation}
    \mathbf{k}(\varphi, \theta) = \frac{2\pi}{\lambda} \Big[\cos{(\theta)}\cos{(\varphi)}, \cos{(\theta)}\sin{(\varphi)}, \sin{(\theta)}\Big]^{\mathrlap{T}}.
    \label{e:wave_vector}
\end{equation}
The narrowband model in~\eqref{eq:RT-based_channel_model} is centered around the frequency $f_c$. It is noted that, for AoA estimation via~\cite{schmidt1986multiple,roy1989esprit}, the modulation of the transmitted signal is irrelevant. For a wideband channel model, we would simply apply the same procedure to each sub-carrier within an Orthogonal Frequency Division Multiplexing (OFDM) block. 
Compared with stochastic channel models, like in~\cite{ma2023reconfigurable,pan2020multicell}, the LoS path is also modeled like the first term in~\eqref{eq:RT-based_channel_model} and the NLoS paths (i.e., the small-scale fading effect) are modeled through a Rayleigh distribution. Thus, both RT-based/geometric and stochastic models result in corresponding mathematical formulations which is useful for replicating and comparing results with other channel models. 

Assuming that $\bm{x}[n]{\in}\mathbb{C}^{N_T}$ is the transmitted signal containing $N_s$ pilot samples, the received signals $\bm{r}_{\text{RIS}}[n]$ and $\bm{r}_{d}[n]$ via the RIS and direct path, respectively, are given by:
\begin{equation}
\begin{aligned}
\bm{r}_{\text{RIS}}[n] \in \mathbb{C}^{N_R} \,&=\, \bm{H_2}[n] \ast \,\big(\boldsymbol{\Phi}\, (\bm{H_1}[n]\,\ast \bm{x}[n])\big), \\
\bm{r}_{d}[n] \in \mathbb{C}^{N_R} \,&=\, \bm{H_d}[n]\,\ast \bm{x}[n], 
\end{aligned}
\end{equation}
where $\ast$ is the convolution operator and $\boldsymbol{\Phi}$ denotes the RIS configuration, specified as a $N{\times} N$ diagonal matrix. It is worth noting that only a single time-domain filter tap is used to model RIS's response, thus, not capturing its frequency selective response~\cite{multiRIS_opt}; this will be addressed in future work. The final received signal at the UE depends on which path arrives first (i.e., the path through RIS or the direct one). The signal arriving second is in fact added to the first appropriately shifted in time. Thus, the total received signal is given by:
\begin{equation}
    \bm{r}_{\text{tot}}=\bm{r}_{\text{RIS}} + \bm{r}_{d} + \bm{w},
    \label{eq:received_signal}
\end{equation}
where $\bm{w}{\in}\mathbb{C}^{N_R}$ is the Additive White Gaussian Noise (AWGN) and $\bm{r}_{\text{RIS}}$ and $\bm{r}_{d}$ are appropriately shifted and added.

\section{B5G/6G Indoor Positioning}
\label{s:problem_localization}
The positioning problem in an indoor setup, such as the one in Fig.~\ref{fig:sim_setup1}, requires processing the received signal $\bm{r}_{\text{tot}}$ in order to extract the position coordinates of the UE, based on some reference points (e.g., APs' or RISs' locations). Therefore, by exploiting the new B5G/6G network infrastructure, namely ULA-based UEs and the RIS, the angle and ranging (or Time of Arrival (ToA)) of the LoS RIS-to-UE path can be estimated, leading to accurate positioning. Assuming perfect ToA estimation (RIS-aided ToA estimation is left for future work), 
here we focus on solving the AoA estimation problem. Since super-resolution algorithms~\cite{schmidt1986multiple,roy1989esprit} calculate the AoA of the LoS path with high accuracy when LoS path is stronger than other paths, a configuration that beamforms towards this LoS path is needed. But, since the position of the UE is not known to directly beamform to it using the RIS, an offline codebook of RIS configurations is designed. Subsequently, the one that maximizes UE's  received power is selected. 

The proposed codebook design method tries to shape the power radiation pattern of the RIS in a sense that it tries to find for each desired reflection angle (angles covering the desired space resolution) an RIS configuration that reflects as much power as possible to that particular angle and less power to the others. The power radiation pattern of RIS is given by~\cite{ramezani2023broad}:
\begin{equation}
    \text{A}(\varphi, \theta) = |\bm{\omega}_{\bm{\theta}}^{T}\,(\mathbf{a}_{\text{RIS}}(\varphi_{\text{AoA}},\theta_{\text{AoA}}) \odot \mathbf{a}^{*}_{\text{RIS}}(\varphi, \theta))|^2,
    \label{eq:power_factor}
\end{equation}
where $\odot$ is the Hadamard product, $(\varphi, \theta)$ refers to the angle pair where the RIS reflects power, $(\varphi_{\text{AoA}},\theta_{\text{AoA}})$ is the AoA pair on the RIS, $\bm{\omega}_{\bm{\theta}} = [e^{\text{j}\theta_{1}}, \ldots , e^{\text{j}\theta_{N}}]^\mathsf{T} \in \mathbb{C}^N $ is the RIS configuration, and $\mathbf{a}_{\text{RIS}}(\varphi,\theta)$ is given by~\eqref{e:array_response_vec}. In fact, the RIS array response vector incorporates not only geometric features of the surface, but also the radiation pattern of each RIS element. 
However, 
here an isotropic radiation pattern for each RIS element is assumed. 
Therefore, to beamform to a specific direction/angle pair $(\varphi,\theta)$, we need to maximize the $\text{A}(\varphi,\theta)$ in~\eqref{eq:power_factor} with respect of $\bm{\omega}_{\bm{\theta}}$. Using the maximum ratio transmission solution~\cite{765552}, the $\text{A}(\varphi,\theta)$ is maximized for
\begin{equation}
    \bm{\omega}_{\bm{\theta}}^{T} = (\mathbf{a}_{\text{RIS}}(\varphi_{\text{AoA}},\theta_{\text{AoA}}) \odot \mathbf{a}^{*}_{\text{RIS}}(\varphi,\theta))^{H}.
    \label{opt_config}
\end{equation}
However, the result in~\eqref{opt_config} corresponds to the continuous case, where we have infinite angle resolution for each RIS-element's configuration. For practical RIS implementations, where each element's configuration applies for two discrete values (e.g., $\theta_{i}{\in}\{-\frac{\pi}{2},\frac{\pi}{2}\}, \forall i$), the solution in~\eqref{opt_config} needs to be quantized, leading also to a quantized version of the codebook.

After the codebook generation, the protocol for RIS-aided localization continues with specifying the RIS configuration included in the codebook that maximizes the received power at the Rx (i.e., beam scanning), thus, solving the problem:
\begin{equation} \bm{\Phi}_0 = \underset{\bm{\Phi}\, \in \,\mathcal{CB}}{\operatorname{argmax}} \,\,\,\, \bm{H_2}[n] \ast \,\big(\Phi\, (\bm{H_1}[n]\,\ast \bm{x}[n])\big) + \bm{r}_{d}, 
\label{eq:max_problem}
\end{equation}
where $\mathcal{CB}$ denotes the codebook. The configuration $\bm{\Phi}_0$ (like all RIS phase configurations in $\mathcal{CB} $) is constructed to perform beamforming to a certain angle. Therefore, the angle at which this configuration reflects the signal, is also a first estimate of the angle at which the UE is located. Using this configuration (i.e., $\bm{\Phi}_0$) to improve the power of the LoS path against the NLoS terms when executing MUSIC algorithm~\cite{schmidt1986multiple}, results in enhanced AoA estimation accuracy.

It is noted that, sometimes when running the MUSIC algorithm~\cite{schmidt1986multiple} using the observation $\bm{r}_{\text{tot}}$ in~\eqref{eq:received_signal}, the AP-to-UE channel is stronger than the channel through the RIS~\cite{zheng2022survey}. This is mainly due to the double path-loss present in the virtual, through the RIS, LoS path~\cite{zheng2022survey}. In this case, the ON/OFF protocol of~\cite{zheng2022survey} can be applied that uses the RIS to estimate the measurement received from the direct channel. The $\bm{r}_{d}$ term can be then subtracted from the measurements and MUSIC estimates the LoS-path's AoD from the RIS with high accuracy, since interference from signals arriving directly from the AP is reduced. To estimate the $\bm{r_d}$ term, we sequentially apply the configurations $e^{j\frac{\pi}{2}} {=} j$ ($\boldsymbol{\Phi}^{1}$) and $e^{-j\frac{\pi}{2}} {=} -j$, (i.e., $\boldsymbol{\Phi}^{2}{=}-\boldsymbol{\Phi}^{1}$) to each RIS element, resulting in the measurements:
\begin{equation}
    \begin{aligned}
        \bm{r}_1  &= \bm{H_2}[n] \ast \,\big(\Phi^{1}\, (\bm{H_1}[n]\,\ast \bm{x}[n])\big) + \bm{r}_{d} +\bm{w} \\ 
        \bm{r}_2  &= \bm{H_2}[n] \ast \,\big(-\Phi^{1}\, (\bm{H_1}[n]\,\ast \bm{x}[n])\big) + \bm{r}_{d} + \bm{w}\\
   &= -\bm{H_2}[n] \ast \,\big(\Phi^{1}\, (\bm{H_1}[n]\,\ast \bm{x}[n])\big) + \bm{r}_{d} + \bm{w}.
    \end{aligned}
\end{equation}
Adding $\bm{r}_1$ and $\bm{r}_2$ we obtain $\bm{r}_d^{\text{est}} = (\bm{r}_1 + \bm{r}_2)/2$. Thus, the final UE's observation is $\bm{r}_{\text{tot}} = \bm{r}_{\text{RIS}} + \bm{w}$. Therefore, the measurement obtained after removing  $\bm{r}_d^{\text{est}}$ depends on the RIS configuration, and if it is appropriately selected from $\mathcal{CB}$, AoA estimation's accuracy can be significantly enhanced. 

\section{B5G/6G-based Mapping}
\label{s:problem_mapping}
Radio mapping from~\eqref{eq:RT-based_channel_model} concerns specifying the location of scattering points in space, that also create NLoS paths. In principle, 
NLoS paths of a single reflection point and of which the scattering point is located at the front of RIS are of interest, so that the path can be activated with some RIS configuration. 
The remaining paths are significantly attenuated, and thus, treated as interference. 
Additionally, since ULA-based UEs are commonly used, only NLoS paths with $\theta = 0$ can be resolved. So, henceforth the notation for the $\theta$-angle is dropped.

Continuing with the hypothesis that the LoS and single-reflected NLoS paths are stronger over all others (both as a summation and as individual paths), the algorithmic description of the proposed mapping protocol follows. Having already estimated the position of the UE, the next step is to perform the beam scanning technique again, but this time activating the NLoS paths sequentially. However, the attenuation of the NLoS channel terms is large compared to the LoS term. Therefore, even for RIS configurations in $\mathcal{CB}$ that do not beamform in the direction of the LoS path, but in the direction of one of the NLoS paths, the strength of the LoS term is dominant. This leads MUSIC algorithm in estimating the angle of the LoS path and not that of the NLoS-path. To deal with this phenomenon, an approach similar to the ON/OFF protocol is adopted. However, in this case, the goal is to remove the LoS path's term from the measurements.

\begin{algorithm}[t]
\scriptsize
\KwIn{$\bm{r}_{\text{tot}}$ from~\eqref{eq:received_signal}.}
\KwOut{UE's position:$(\varphi_{\text{AoA}}^{\text{est}},\tau^{\text{est}})$ and scatterer position.}
\nl $\mathcal{CB}$ design based on~\eqref{eq:power_factor} and~\eqref{opt_config}\;
    \nl Solve problem~\eqref{eq:max_problem} and transmit using the $\bm{\Phi}_0$ from~\eqref{eq:max_problem}\;
    \nl Execute the ON/OFF protocol and MUSIC~\cite{schmidt1986multiple} to obtain $\varphi_{\text{AoA}}^{\text{LoS}}$\;
    \nl Specify $\tau^{\text{est}}$ using an UWB equipment\;
    \nl \For {\texttt{$\bm{\Phi} \in \mathcal{CB}$}}{
            \nl Use the ON/OFF protocol and~\eqref{eq:LoS_term},~\eqref{eq:mapping_measurement} to generate the final observation\;
            \nl Execute MUSIC~\cite{schmidt1986multiple} to obtain $\varphi_{\text{AoA}}^{\text{NLoS}}$\;
            \nl \If{$\varphi_{\text{AoA}}^{\text{NLoS}} \neq \varphi_{\text{AoA}}^{\text{LoS}}$}{
                \nl Using geometry manipulations, find a scattering-object's position\;
            } 
        }
\caption{{\bf Positioning and Mapping Protocol}
\label{algorithm}}
\end{algorithm}

Denoting by $\varphi_{\text{AoA}}^{\text{est}}$, $r^{\text{est}}$, and $\tau^{\text{est}} = \frac{r^{\text{est}}}{c}$ ($c$ is the speed of light), the azimuth AoA estimation at the UE, the ranging estimation, and the ToA estimation for the LoS path between the RIS and UE respectively, the LoS term of the RIS-to-UE channel from~\eqref{eq:RT-based_channel_model} can be estimated using the expression:
\begin{equation}
    \bm{H}_{2\,\,\text{LoS}}^{\text{est}}= h_2^{\text{loss}}e^{j\phi_2}\bm{\mathbf{a}}_{\text{UE}}(\varphi_{\text{AoA}}^{\text{est}})\bm{\mathbf{a}}_{\text{RIS}}^{H}(\varphi_{\text{AoD}}^{\text{est}}) \in \mathbb{C}^{N_R \times N},
    \label{eq:LoS_term}
\end{equation}
where $h_2^{\text{loss}} = 20\operatorname{log}_{10}\big(\frac{4\pi r^{\text{est}}}{\lambda}\big)$ and $\lambda$ is the carrier frequency wavelength. This expression assumes that the target is in the Tx's far field. Moreover, $\phi_2 = \operatorname{mod}(-2\pi f_c \tau^{\text{est}},2\pi)$ and $\varphi_{\text{AoD}}^{\text{est}}$ follows from $\varphi_{\text{AoA}}^{\text{est}}$ via geometric manipulations. In exactly the same way, the term $\bm{H}_{1\,\,\text{LoS}}^{\text{est}} {\in} \mathbb{C}^{N{\times}N_T}$ can be constructed, which refers to the AP-to-RIS channel matrix, with the difference that the geometry of the AP-to-RIS channel is already known, and thus, the various parameters needed for $\bm{H}_{1\,\,\text{LoS}}^{\text{est}}$ computation are not estimates, but exact values. Whenever a measurement is received at the UE via a known RIS configuration, the measurement $\bm{r}_{\text{LoS}}[n] {\in} \mathbb{C}^{N_R} = \bm{H}_{2\,\,\text{LoS}}^{\text{est}} \, \bm{\Phi}\, \bm{H}_{1\,\,\text{LoS}}^{\text{est}}\,\bm{x}[n]$ can be constructed and subtracted from the received signal:
\begin{equation}
\bm{r}_{\text{NLoS}}[n] = \bm{r}_{\text{tot}}[n] - \bm{r}_{\text{LoS}}[n-\lfloor{\tau^{\text{est}}F_s\rfloor}],
\label{eq:mapping_measurement}
\end{equation}
where $F_s$ is the UE's sampling frequency and $\lfloor{x\rfloor}$ denotes the integer part of $x$. So, the impact of the strong LoS term is removed from the measurements. To this end, the beam sweeping technique is used again, but now the measurement on which MUSIC is executed, is the $\bm{r}_{\text{NLoS}}[n]$ in~\eqref{eq:mapping_measurement}. Again, due to the fact that it is not possible to perfectly eliminate the LoS term, for most configurations of the codebook, the LoS path's angle is given as a solution. However, when targeting at an NLoS path using the RIS, the angle found by MUSIC is that of this NLoS path. Thus, the location of the scattering point can be specified as the intersection of the two straight lines, the one that starts from the RIS and leaves at the angle at which RIS beamforms and the straight line that arrives at the UE and coincides with the AoA calculated by MUSIC. This process is repeated for each configuration that enables a different NLoS path, i.e., MUSIC gives other AoA estimation than the LoS path's AoA. To this end, all possible scattering objects can be positioned in an indoor place, thus providing a mapping of the space. The overall protocol for joint positioning and mapping is summarized in Algorithm~\ref{algorithm}.

\section{Simulation Results}
\label{s:simulations}

\begin{figure}[t]
\centering
\begin{minipage}{0.25\textwidth}
  \centering
\includegraphics[width=0.78\textwidth]{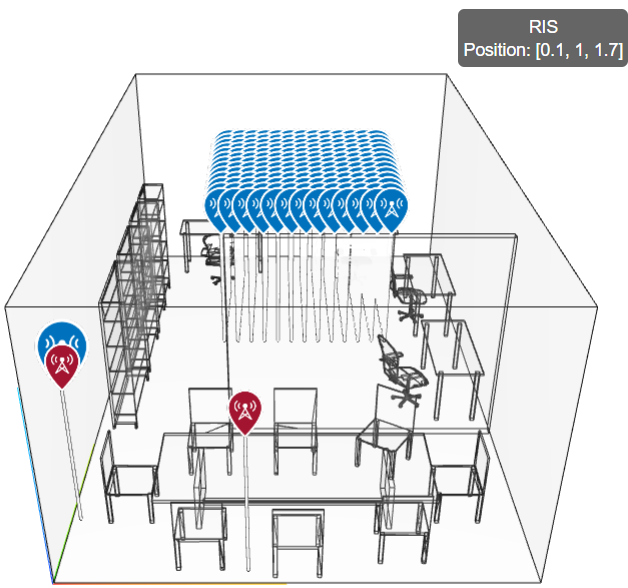}
\subcaption[first caption.]{Test Rx positions of the UEs.}\label{fig:1a}
\end{minipage}%
\begin{minipage}{0.25\textwidth}
  \centering
\includegraphics[width=0.92\textwidth]{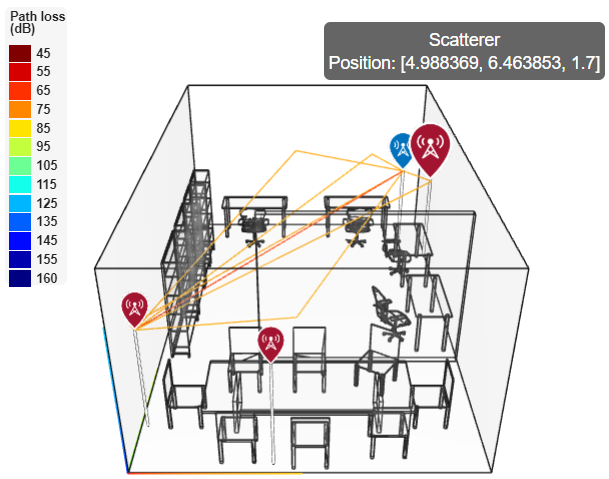}
\subcaption[second caption.]{Scatterer positions for mapping.}\label{fig:1b}
\end{minipage}%
\caption{The simulated  indoor environment for localization and mapping.} \label{fig:exp_loc_setup}
\end{figure}

The experimental setup used to assess the proposed protocol for RIS-aided localization and mapping is illustrated in Fig.~\ref{fig:exp_loc_setup}, where a $32{\times}32$ URA RIS operating at $f_c{=}3.5$~GHz is utilized. All nodes are on the same azimuth plane and to validate the proposed method's accuracy, we have run Algorithm~\ref{algorithm} for each UE position in Fig.~\ref{fig:1a} and averaged the results. The test UE positions were determined by the arrangement depicted in Fig.~\ref{fig:1a}, where UEs were located in a rectangle determined by $x {\in} [1.5,3.5]$ (abscissa) and $y {\in} [5,7]$ (ordinate). The channel model followed a geometric RT-based approach and the setup entailed a multi-antenna Tx directly beamforming towards the RIS (simulating a highly directive horn antenna) and multi-antenna UEs with fixed antenna orientation at the $x$-axis.

\subsection{Positioning}
\begin{table}[tb]
\caption{Evaluation of Algorithm~\ref{algorithm}.\vspace*{-8pt}}
\begin{center}
\scalebox{0.92}
{
\setlength{\tabcolsep}{2pt}
\begin{NiceTabular}{c@{\hskip 6pt}cc
>{\centering\arraybackslash}p{1cm}
>{\centering\arraybackslash}p{1cm}S[table-format=3.2]}
\CodeBefore
\rowcolor{teal!10}{1}
\Body

     \toprule
     Phase shifters & RIS configuration & Method & {Peak} & {Average}  & {Variance} \\
     \midrule
     Continuous  &$\bm{\omega}_{\bm{\theta}}^{T}$ from~\eqref{opt_config} & BeS$^\dagger$ & 12.62\degree  & 1.03\degree  & 2.28\degree \\
     \midrule
     1-bit RIS  &  Quantized $\bm{\omega}_{\bm{\theta}}$ & BeS & 58.28\degree & 9.93\degree &280.32\degree  \\
     1-bit RIS  &  Quantized $\bm{\omega}_{\bm{\theta}}$ & BeS \& MUSIC & 10.68\degree & 0.86\degree &1.82\degree\\
     \bottomrule
     \multicolumn{6}{l}{$^\dagger$ BeS: Beam Sweeping}
\end{NiceTabular}
}
\label{Table:beamsweeping_algorithms}
\end{center}
\vspace{-0.1cm}
\end{table}

Table~\ref{Table:beamsweeping_algorithms} presents statistical measures for the absolute estimation error $\tilde{e} = \big| \varphi_{\text{AoD}} - \varphi_{\text{AoD}}^{\text{est}} \big|$, 
where $\varphi_{\text{AoD}}$ denotes the actual azimuth angle between the RIS and the UE. This table also contains the maximum, mean, and variance of the estimation error for each method. The methods analyzed encompass the continuous solution for the codebook design and quantization of the continuous solution employing $1$-bit phase shifters, to be consistent with practical RIS implementations. Furthermore, the performance of the $1$-bit codebook was assessed in conjunction with the MUSIC algorithm. Finally, for the reproducibility of the results, it is mentioned that the codebook described in section~\ref{s:problem_localization} is constructed with a resolution of approximately two degrees. 

It can be observed from Table~\ref{Table:beamsweeping_algorithms} that, even with $1$-bit RIS codebook obtained by quantizing the continuous solution, a satisfactory solution can be achieved, even in the challenging case where each individual channel is a strong multipath. However, side lobe effects~\cite{liu2024quantization} occurring for $1$-bit RISs, can activate other NLoS paths, and thus, both quantized techniques sometimes lead to misleading results; the large peak error in the $1$-bit beam sweeping technique results from quantization of the continuous solution of a non-convex problem.

\subsection{Mapping}
\begin{table}[tb]
\caption{RIS-UE paths obtained by ray tracing.\vspace{-0.2cm}}
\label{t:paths_mapping}
\vspace{1ex}
\centering
\scalebox{0.9}{
\begin{NiceTabular}{ccS[table-format=3.2]S}
\CodeBefore
\rowcolor{teal!10}{1}
\Body
\toprule
Path Number &Path Type  & {Azimuth  AoA} & {Elevation AoA (deg, \degree)}\\
\midrule
1. &LoS &53.75\degree &0\\
2. &NLoS &127.36\degree & 8.56e-14\\
3. &NLoS&53.75\degree&21.96\\
4. &NLoS&47.91\degree&0\\
5. &NLoS&53.75\degree&-24.56\\
6. &NLoS&61.29\degree&0\\
\bottomrule
\end{NiceTabular}}
\vspace{3pt}
\end{table}
As shown in Fig.~\ref{fig:sim_setup1}, there are many NLoS paths that need to be resolved in this setup. Table~\ref{t:paths_mapping} lists all RIS-to-UE paths whose reflection-point coordinates need to be specified by Algorithm~\ref{algorithm}. Since not all of these paths are resolvable with a ULA-based Rx, in this preliminary analysis for mapping, simulation results for resolving the single-reflected NLoS paths with $\theta_{\text{AoA}}{\approx} 0$ are presented, of which scattering point is not behind the RIS. Moreover, since MUSIC spectrum is periodic in the interval $[-180\degree,180\degree]$, we had to limit it to the interval $[0\degree,180\degree]$, thus, paths incident on the backside of the Rx-ULA were not resolved. For this reason, we assumed that the Rx-UE was equipped with a patch antenna, and thus, could not receive from its back side. Therefore, mapping only concerns the space in front of the UE. Mapping the back space requires the UE to rotate facing the back space. Therefore, in this work, we focused on resolving only the NLoS path $4$.  

As can be seen from Fig.~\ref{fig:1b}, the calculation of the scattering point is very close to the actual reflection point that creates this particular NLoS path. 
If we now run the Algorithm~\ref{algorithm} for each UE position in Fig.~\ref{fig:1a}, the average value of the estimation error for the scattering point position is $0.21$~m, with a standard deviation of $0.23$~m, indicating that the RIS can enable nearly dm-level of mapping accuracy.   

\section{Conclusion}
\label{s:conclusion}
This paper presented a comprehensive framework that leverages beam sweeping through codebook-based beamforming at the RIS, signal component extraction, and a super-resolution algorithm to enhance user localization and mapping in future 6G wireless networks. By addressing the challenges of indoor environments with multipath effects, especially using an $1$-bit RIS, and validating with accurate 3D RT models, we demonstrated the robustness and effectiveness of our proposed approach in practical indoor scenarios.

\section{Acknowledgement}
This work has been supported by the ESA Project PRISM: RIS-enabled Positioning and Mapping (NAVISP-EL1-063). 

\balance
\bibliographystyle{IEEEtran}
\bibliography{IEEEfull,irs}

\end{document}